\documentclass[12pt]{article}
\usepackage[T1]{fontenc}
\usepackage{graphicx}
\usepackage{subcaption}
\usepackage{amsmath}
\usepackage{amssymb}
\usepackage{algorithm}
\usepackage{algpseudocode}
\usepackage{hyperref}
\usepackage[skip=10pt]{caption} % for \captionsetup[table]{skip=10pt}

\title{Improving Community Detection in Academic Networks by Handling Publication Bias \thanks{This paper is an extended version of a work accepted at ASONAM 2025.}}
\author{%
  Md Asaduzzaman Noor\thanks{\texttt{mdasaduzzamannoor@montana.edu}} \and
  John Sheppard\thanks{\texttt{john.sheppard@montana.edu}} \and
  Jason Clark\thanks{\texttt{jaclark@montana.edu}} \\
  Montana State University, Bozeman MT-59717, USA
}
\date{}

\begin{document}
\maketitle

\begin{abstract}
Finding potential research collaborators is a challenging task, especially in today’s fast-growing and interdisciplinary research landscape. 
While traditional methods often rely on observable relationships such as co-authorships and citations to construct the research network, in this work, we focus solely on publication content to build a topic-based research network using BERTopic with a fine-tuned SciBERT model that connects and recommends researchers across disciplines based on shared topical interests.
A major challenge we address is publication imbalance, where some researchers publish much more than others, often across several topics. 
Without careful handling, their less frequent interests are hidden under dominant topics, limiting the network’s ability to detect their full research scope. To tackle this, we introduce a cloning strategy that clusters a researcher’s publications and treats each cluster as a separate node. 
This allows researchers to be part of multiple communities, improving the detection of interdisciplinary links. 
Evaluation on the proposed method shows that the cloned network structure leads to more meaningful communities and uncovers a broader set of collaboration opportunities.

\end{abstract}

\section{Introduction}

Research collaboration plays a crucial role in advancing scientific discovery, often leading to more impactful and interdisciplinary outcomes. 
As the volume of scholarly publications continues to grow, recommending meaningful collaborations has become an increasingly challenging problem in the research community. 
Most existing approaches rely on observable relationships, such as co-authorship or citation networks, to recommend collaborators. 
While effective to some extent, these methods tend to reinforce known connections, missing opportunities to connect researchers based on shared topical interests.

We argue that topical similarity, which can be derived from publication content, is a powerful relation for identifying potential collaborations. 
Two researchers working on similar themes may never have co-authored a paper or even be aware of each other’s work. 
By looking at what researchers are actually publishing, we can uncover hidden connections and make collaboration recommendations that are more diverse, going beyond the usual disciplinary lines.

Although prior work has leveraged publication data, most focus narrowly on ranking collaboration candidates based on content similarity scores. 
These models typically provide limited interpretability, offering little beyond a top-$k$ list with abstract relevance scores. 
In contrast, social network analysis (SNA) offers a broader perspective. 
Not only does it enable content-based recommendations, but it also reveals how researchers are organized around shared interests, highlights influential nodes, and offers more interpretable, community-driven insights through network structures and visualizations.

In this work, we construct a scholarly network entirely based on topic similarity, using basic metadata (publication titles and abstracts) to group researchers by shared research themes. 
This builds on our prior work \cite{noor_scholarnode,noor_finding}, where we demonstrated that content-based networks can enhance the novelty and diversity of collaboration recommendations. 
However, one key challenge we address in this paper is possible imbalance in publication counts across researchers. 
High-output researchers often work on multiple topics, but their less frequent interests are overshadowed by their dominant ones when building similarity-based networks. 
As a result, meaningful connections between those smaller topics can be missed.

To address this, we introduce a cloning strategy for highly productive researchers. 
By clustering their publications into coherent topical groups, we create multiple ``clones” that  represent distinct research areas. 
This allows these researchers to participate in multiple communities, improving the detection of diverse and interdisciplinary collaborations. 
%To our knowledge, this is the first effort to address publication imbalance explicitly in topic-based community detection for research collaboration.
Our key contributions are:
\begin{itemize}
    \item We propose a cloning-based strategy to handle publication imbalance, enabling more accurate and diverse community detection.
    \item We construct a topic-based researcher similarity matrix with clones using BERTopic applied to publication titles and abstracts.
    \item We evaluate the quality of the resulting communities empirically and demonstrate the benefits of our approach for collaboration recommendations.
\end{itemize}

\section{Related Work}
One of the key goals of researcher social network analysis is to provide potential research collaboration recommendations. 
Most existing work in this area relies heavily on direct relationships between researchers, such as co-authorship or citation links, to suggest collaborators or identify research communities. 
While effective to some extent, these approaches often overlook the topical diversity of individual researchers, limiting the ability to connect them based on shared research interests.

Earlier work often considered collaboration recommendation as a link prediction problem, where the goal is to infer new or future connections within the network. 
For instance, Nowell et al. \cite{liben2003link} constructed co-authorship networks and proposed prediction methods based on vertex neighborhoods and network path ensembles. 
Later, Backstrom et al. \cite{backstrom2011supervised} introduced supervised random walks on co-authorship networks to improve link prediction.

Several hybrid approaches were also proposed that combined direct relationships with content-based features to improve recommendation quality. 
Yang et al. \cite{yang_2015} integrated research expertise, co-authorship, and institutional ties, using a language model for semantic similarity and an SVM-Rank framework for matching. 
Kong et al. \cite{kong2016exploiting} built a co-authorship network, applied Word2Vec to publication titles, and used clustering and cosine similarity within a Random Walk model. 
Zhou et al. \cite{zhou_2018} created a multidimensional academic network using ResearchGate data and introduced time-aware edge weights, favoring recent collaborations with an improved Random Walk algorithm.
All of the above hybrid methods rely on some form of direct link to construct the network, which limits their ability to identify entirely new, interdisciplinary collaborations. 
They tend to reinforce existing connections rather than explore potential new connections.
%novel ones based purely on topical alignment.

Other work has used only publication data to drive recommendations. 
Liang et al. \cite{liang2017recommendation} used LDA to model research fields and compared topic vectors to suggest cross-disciplinary collaborations. 
Kong et al. \cite{kong2017exploring} proposed the BCR model, incorporating evolving interests and academic influence via time-weighted topic distributions and cosine similarity.

Existing content-based approaches often focus on document similarity to generate top-$k$ recommendations. 
However, we argue that such models can lack interpretability, providing only similarity scores without much explanation. 
By integrating social network analysis into content-based data, we aim to make the recommendation process more transparent and informative. 
Community detection, in particular, goes beyond isolated suggestions, offering a structural view of researcher clusters, topic distributions, and key participants.
%how researchers cluster, how topics are distributed, and who plays a central role.

In this work, we construct a research network using topic-based similarity between researchers while accounting for publication count imbalance to ensure fairer and more meaningful community detection. 
%To our knowledge, this is the first effort to explicitly consider publication imbalance in identifying topical research communities.
To our knowledge, this is the first effort to address publication imbalance explicitly in topic-based community detection for research collaboration.

\section{Dataset}
To build the researcher dataset, we used Montana State University's (MSU) current faculty list and retrieved their publication history using OpenAlex \cite{priem2022openalex}, an open source API for accessing scholarly metadata. We collected publication titles, abstracts, and author IDs from 2004 to 2025 for papers affiliated with MSU faculty.
In total, we extracted metadata for $9,768$ publications. 
To ensure meaningful topic distributions for our network construction, we excluded researchers with fewer than five publications. 
This resulted in a final dataset of $296$ faculty members, with a maximum of $190$ publications for a single researcher and mean and median publication counts of $33$ and $22$, respectively.

\begin{figure}[t]
    \centering
    \includegraphics[width=0.8\linewidth]{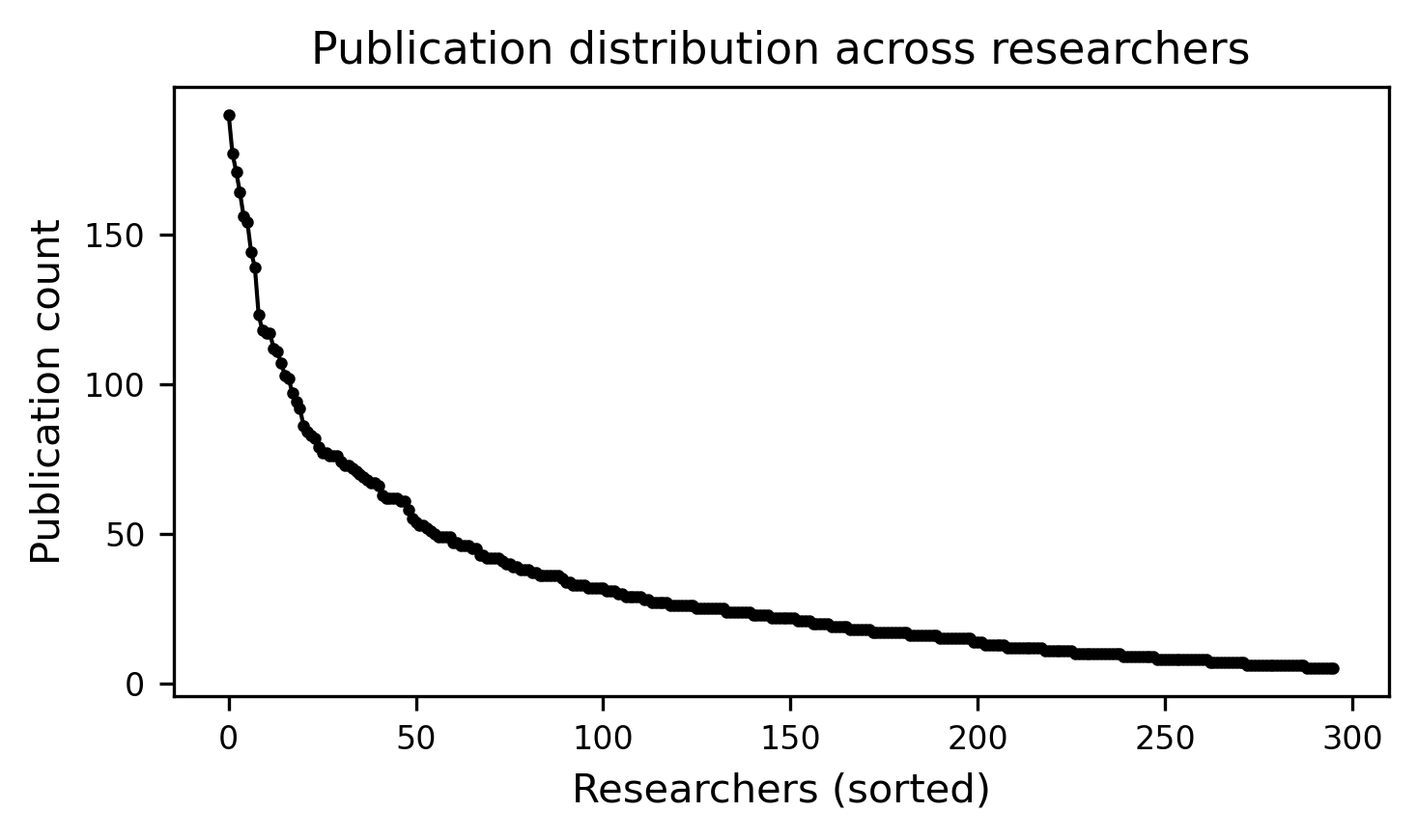}
    \caption{Researchers' publication count in descending order}
    \label{fig:pub_count}
\end{figure}

Figure \ref{fig:pub_count} shows the publication count distribution across the researchers in descending order. 
The distribution is heavily right-skewed, with a small number of researchers publishing significantly more than others. 
This imbalance neccessitated our addressing publication count disparities for effective community detection.
%in the topic-based network.

\section{Methodology}
\begin{figure}[t!]
    \centering
    \includegraphics[width=1\linewidth]{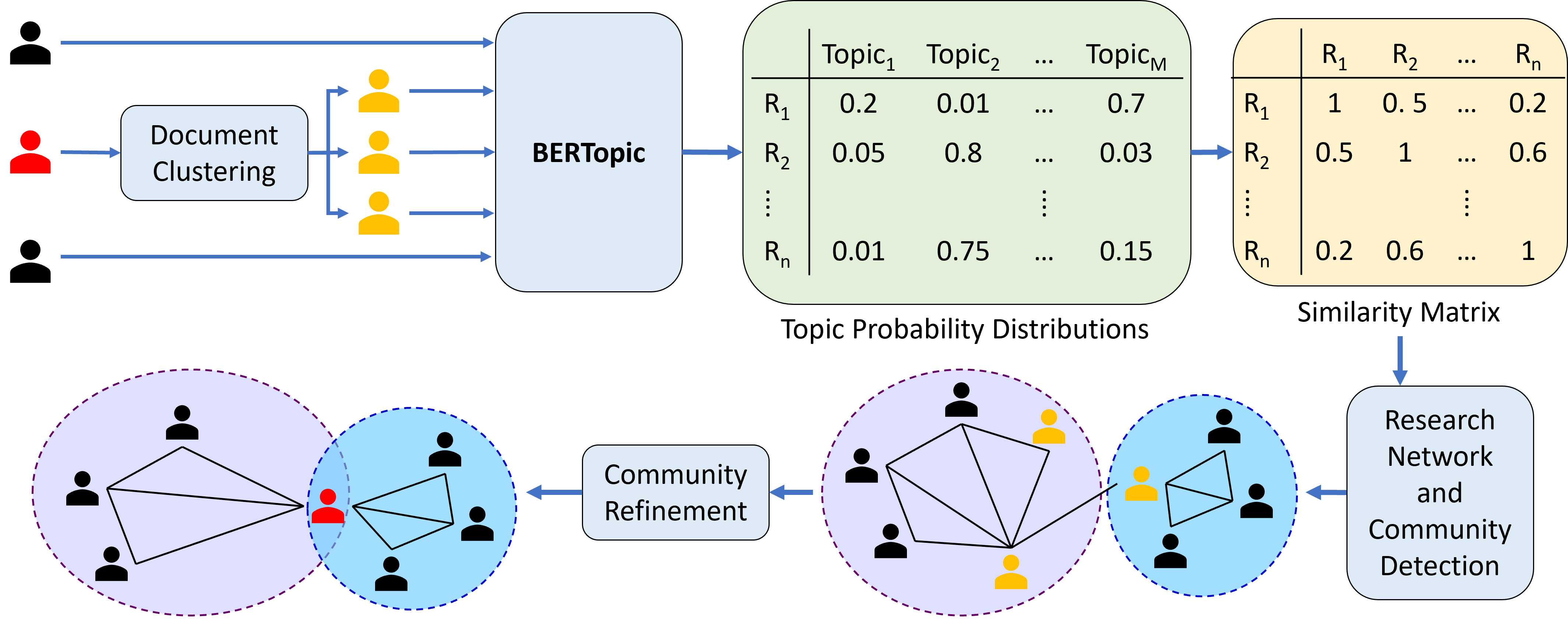}
    \caption{Overview of the proposed methodology}
    \label{fig:outline}
\end{figure}

Our proposed method is depicted in Figure \ref{fig:outline}. 
First, we train a topic model using the publication metadata of all researchers. 
Next, we create clones for high-publication researchers by clustering their publications. 
Then, we compute the topic probability distributions for each researcher (including the clones) and use these to build a topic similarity matrix between researcher pairs. 
From this matrix, we construct a research network and apply community detection. 
Finally, we refine the detected communities by merging duplicate nodes (i.e., clones) to obtain the final set of research communities that share similar topical interests. 
We provide a detailed explanation of each step in the following sections.

\subsection{Topic Modeling}
The first step in creating a collaborative research network involves building a topic model from the researchers' publications. 
While any topic modeling technique could be used for this step, we chose BERTopic \cite{grootendorst2022bertopic}, which extracts topics from text while emphasizing contextual information.
%The process begins with the use of a pre-trained sentence transformer model (e.g., BERT \cite{bert_mlm}) to generate document embeddings in a high-dimensional space. 
%Next, a dimensionality reduction technique (e.g., UMAP \cite{McInnes2018}) is applied to these embeddings to manage the high-dimensional data. 
%Following this, a clustering algorithm (e.g., HDBSCAN \cite{McInnes2017}) groups the documents based on similarity, with each cluster representing a distinct topic. 
Similar to Latent Dirichlet Allocation (LDA) \cite{lda}, which generates a word probability distribution for each topic, BERTopic uses a class-based tf-idf technique to identify the top words associated with each topic. 
Additionally, BERTopic allows for the extraction of topic probability distributions for each document, reflecting the degree to which each document belongs to a given topic.

%For our analysis, w
We used the publications of all researchers in our dataset as the corpus. 
Previous studies have shown that fine-tuning the sentence-embedding model on a specific dataset can enhance the performance of the transformer model \cite{ding2023parameter,gururangan2020,tinn2023fine}. 
Therefore, we fine-tuned the sentence transformer with our corpus using Masked Language Modeling (MLM) \cite{bert_mlm}.
In MLM, words or subwords in a sentence are replaced randomly with a special token (e.g., [MASK]). 
The model’s objective is to predict these masked tokens based on the context provided by the surrounding unmasked words. 
While pre-trained sentence transformers are trained on large datasets, applying MLM to our specific corpus enables the model to better capture context relevant to our research domain.
From the fine-tuned BERTopic model, we obtain topics, topic-word distributions, and document-topic probability distributions, which are used in the subsequent steps of our methodology.

\subsection{Cloning High-Impact Researchers via Publication Clustering}
It is common for research institutions to have researchers with a high publication count, in contrast to others with fewer publications. 
When extracting topic probability distributions using all publications of a high-count researcher, the distribution often favors topics that appear frequently across most of their work. 
This can lead to missing less frequent but still relevant topics. 
Since the goal is to identify research communities based on topic similarity, high-publication researchers may end up belonging to a single community aligned with only their dominant topical interests.

A key characteristic of many high-publication researchers is that they often work across multiple diverse areas. As such, their less frequent topics should also connect them to other communities beyond their main focus. To capture this, we introduce a cloning strategy that allows a researcher to be represented in multiple communities.

In our dataset, the median publication count per researcher is $22$. Based on this, we define high-impact researchers as those with more than $1.5$ times the median (i.e., over $33$ publications) and select them for cloning. 
The clones are generated by clustering documents based on topic similarity.
Specifically, we first collect all papers of a high-impact researcher and represent them using the fine-tuned sentence embedding model trained earlier for global topic modeling. 
Consistent with BERTopic we then apply UMAP for dimensionality reduction, followed by HDBSCAN to group the publications based on similarity. 
%HDBSCAN is well-suited here as it does not require the number of clusters to be predefined.

The number of clusters produced by HDBSCAN determines how many clones are created. 
For instance, if researcher A's publications form three clusters, three clones, A1, A2, and A3, are created, each containing the documents from the corresponding cluster. 
Note that HDBSCAN may assign a label of $-1$ to documents it identifies as outliers or noise. Since we are clustering publications, we include these as a separate clone as they still provide useful information about the researcher's broader topical activity.

\subsection{Researchers' Topic Similarity Matrix}

Following the cloning step, which increases the total number of researchers by adding clones for high-publication individuals, the next step involves computing the topic probability distribution for each researcher. 
The fine-tuned BERTopic model provides topic probability distributions at the document level. 
Briefly, BERTopic calculates the topic distribution for a document by dividing it into chunks of words or token sets. 
It then applies a sliding window to assign topics to each chunk. 
The overall topic distribution of a document is derived by summing the topic assignments of all chunks and normalizing the resulting vector.

To compute the topic distribution for a researcher, we aggregate the topic distributions of all their publications and normalize the sum. 
Specficially, let $\mathbf{N}$ be the total set of publications, and $\mathbf{N}_i \subseteq \mathbf{N}$ denote the set of publications authored by researcher $i$, with $n_i = |\mathbf{N}_i|$. Let $p_j \in \mathbf{N}_i$ represent an individual publication of researcher $i$, and let $\boldsymbol{\theta}_{p_j}$ be the topic distribution of publication $p_j$. 
The topic distribution for researcher $i$, denoted as $\boldsymbol{\Theta}_i$, is calculated as:
\[
\boldsymbol{\Theta}_i = \frac{1}{n_i} \sum_{p_j \in N_i} \boldsymbol{\theta}_{p_j}
\]

After computing topic distributions for all researchers, we construct the researchers' topic similarity matrix. 
Since BERTopic provides probability distributions, we employ the Jensen-Shannon Divergence (JSD) \cite{Ross97} to quantify the similarity between two researchers' topic profiles, where 
%JSD is a symmetric variant of the Kullback-Leibler (KL) Divergence and is suitable for comparing probability distributions. For two distributions $P$ and $Q$, JSD is defined as:
\[
\text{JSD}(P \| Q) = \frac{1}{2} D_{\text{KL}}(P \| M) + \frac{1}{2} D_{\text{KL}}(Q \| M).
\]
Here $D_{\text{KL}}$ denotes the KL divergence, and $M = \frac{1}{2}(P + Q)$ is the mean distribution over $P$ and $Q$. 
The JSD value lies between $0$ and $1$, with lower values indicating higher similarity. For a dataset with $R$ researchers, this results in a topic similarity matrix $\mathcal{S}$ of size $\mathbb{R}^{R \times R}$, where entries correspond to $1-\text{JSD}(\boldsymbol{\Theta}_i,\boldsymbol{\Theta}_j)$.

\subsection{Constructing the Research Network}

A network is represented by a graph structure \( \mathcal{G}(\mathbf{V}, \mathbf{E}) \), where \( \mathbf{V} \) denotes the set of vertices and \( \mathbf{E} \) represents the set of edges connecting the vertices, indicating some form of relationship between them. 
In our case, the researchers are represented by the vertices, and the topic similarity between them reflects the strength of the relationship. 
An adjacency matrix \( \mathbf{A} \) is used to represent the graph. We define the adjacency matrix corresponds to the topic similarity matrix \( \mathcal{S} \).
%, where the weights between vertices are the \( 1 - \text{JSD} \) values. 
%This choice follows the convention that a higher edge weight in a graph signifies a stronger relationship between vertices. Therefore, the element in the adjacency matrix is defined as:

Constructing the network using the similarity matrix results in a fully connected network. 
However, extracting communities from a fully connected network is often challenging because community intuition is based on having many connections within a group, compared to connections outside of it. 
To address this, we also perform edge pruning by removing edges with weights below a certain threshold.

\subsection{Community Detection}

Once the research network is constructed, the next step is to detect communities within the network. 
As our network is based on topic similarity between researchers, the formed communities should ideally group researchers with similar topical interests.

Several algorithms for detecting communities in a weighted, undirected graph exist in the literature \cite{blondel2008fast,ng2002spectral,rosvall2008maps}. 
For our analysis, we utilize the Nested Hierarchical Louvain (NH-Louvain) \cite{noor_nhl} algorithm, which is capable of detecting communities at different hierarchical levels. 
The rationale behind selecting a hierarchical community detection algorithm is that topics often exhibit a natural hierarchical structure. 
For instance, topics like machine learning should be grouped under a broader topic like Artificial Intelligence, which itself falls under even broader categories such as Computer Science, Mathematics, and Statistics. Hence, communities based on topic similarity should also form a hierarchy, with granularity varying according to the level of topical specificity.

The NH-Louvain algorithm discovers communities through the following process. 
The core algorithm, the Louvain method, detects communities by greedily optimizing the modularity score. 
Modularity measures the difference between the fraction of edges within a community and the fraction of edges expected in a random graph with the same degree distribution. 
The Louvain algorithm places vertices in a community as long as doing so increases the modularity score. 
NH-Louvain detects hierarchical communities by first applying the Louvain method to the entire network, then recursively applying Louvain to the communities identified in the previous step, until a stopping criterion is met. 
In our case, we used a minimum community size as a stopping criterion. For a more detailed explanation of the NH-Louvain, please refer to \cite{noor_nhl}.

\subsection{Refining Community Structure} \label{sec:refine_comm}
During the construction of the research network, we introduced clones for high publication-count researchers to account for the diversity of their topical interests. 
Consequently, the same researcher may appear multiple times within a detected community due to topical similarity among their clones. 
Since these clones represent the same underlying researcher, we perform a community refinement step to merge all clones into a single vertex.

This refinement is applied individually to each community, rather than to the full graph. 
The rationale is that, if the clones of a researcher appear in different communities, this separation is intentional as it reflects the researcher's engagement in multiple, distinct topical areas. 
However, if the clones are placed within the same community, they should be merged to represent a single entity.

The refinement process is outlined in Algorithm \ref{alg:refine}. 
The algorithm takes as input the subgraph of a community and its node set, and outputs a new community graph with merged clones. 
First, nodes are grouped by their base identity, assigning a common identifier to all clones of the same researcher. 
Then, for each node, we add its base identity to the graph and iterate over its neighbors. 
If the neighbor has a different base identity, we add or update the edge between their corresponding base nodes in the graph, retaining the maximum observed weight. 
The final refined graph is returned after all nodes are processed.

\begin{algorithm}[t]
\caption{\textsc{RefineCommunity}}
\label{alg:refine}
\begin{algorithmic}[1]
\Require Community subgraph $G_C$, community nodes $C$
\State Group nodes in $C$ by base identity \Comment{Handle clones as one}
\State Initialize new community graph $G'$
\ForAll{node $u$ in $C$}
    \State Add base identity of $u$ to $G'$
    \ForAll{neighbor $v$ of $u$ in $G_C$}
        \State Get base identity of $v$
        \If{base identities of $u$ and $v$ differ}
            \State Add base identity of $v$ to $G'$ if not exists
            \State Add or update edge $(u,v)$ in $G'$ with max weight
        \EndIf
    \EndFor
\EndFor
\State \Return $G'$
\end{algorithmic}
\end{algorithm}

\section{Experimental Design}

In this section, we describe the hyperparameter tuning process for the BERTopic model and how we clustered publications of high-output researchers to generate clones.
To begin, we fine-tuned the sentence embeddings used in BERTopic to better align with our research domain. We started with Hugging Face’s \texttt{allenai/} \texttt{scibert\_scivocab\_uncased} \cite{beltagy-etal-2019-scibert}, a BERT-based model pre-trained on Semantic Scholar papers. 
To adapt it to our dataset, we applied MLM \cite{bert_mlm} on our corpus, masking 15\% of the tokens and training for 40 epochs.

For dimensionality reduction in BERTopic, we used UMAP with the following parameters: \texttt{n\_neighbors = 15}, \texttt{n\_components = 5}, and \texttt{min\_dist = 0.0}. 
Topic clustering was performed using HDBSCAN with \texttt{min\_cluster\_size = 8} and \texttt{min\_samples = 4}, resulting in 445 topics after training with the fine-tuned SciBERT.
To create clones, we embedded the publications of each high-output researcher using the same fine-tuned SciBERT model. 
We then clustered each researcher's documents using HDBSCAN with \texttt{min\_cluster\_size = 10} and \texttt{min\_samples = 5} to form clone groups.
For the class-based TF-IDF representation, we applied standard NLP preprocessing steps, including stopword removal, digit and punctuation filtering, and lemmatization.

Note that hyperparameters were tuned through random search rather than exhaustive search, aiming for interpretable, stable clusters that reflected meaningful topical groupings.

\section{Results and Discussion}

We begin our results discussion focusing on whether cloning improves community detection among researchers—particularly high-impact ones. 
The underlying hypothesis is that the topic distributions of high-impact researchers may be underrepresented, as their diverse research interests can be overshadowed by dominant topics. 
By cloning these researchers through clustering their publications, we aim to reveal more accurate (and possibly overlapping) communities and support more diverse collaboration recommendations. 
Although our method is unsupervised and lacks ground truth for direct comparison, we provide both quantitative and qualitative insights to assess its performance.

In Table \ref{tab:clone_stat}, we present the statistics of the clones for high-impact researchers. 
Our dataset initially included 296 researchers, of which 96 were considered high-impact due to having more than 33 publications (1.5 times 22 median publications). 
These 96 high-impact researchers were then passed to the clustering algorithm (i.e., HDBSCAN) to form groups based on publication similarity. Of these 96 researchers, 68 were assigned to more than one cluster. 
For the remaining researchers, HDBSCAN returned a single cluster, indicating that either their publications were very similar or very diverse (with a cluster level of $-1$), preventing the formation of multiple groups. 
Among the 68 researchers, the maximum number of clones was 10, for a researcher with 173 total publications. 
The median number of publications for the cloned researchers was 23, which is closely aligned with the original median of 22 before cloning.

\begin{table}[t]
\centering
\caption{Summary statistics of the researchers with clones}
\label{tab:clone_stat}
\begin{tabular}{|l|c|}
\hline
Total Number of Researchers                   & 296 \\ \hline
High-Impact Researchers (More than 33 Papers)  & 96  \\ \hline
Researchers with Clones                        & 68  \\ \hline
Max clones for a researcher (173 publications) & 10 \\ \hline
Min clones for a researcher & 2 \\ \hline
Median clones (high-impact researchers) & 3 \\ \hline
Max publications (cloned researchers) & 118 \\ \hline
Avg. publications (cloned researchers) & 27.09 \\ \hline
Median publications (cloned researchers) & 23 \\ \hline
\end{tabular}
\end{table}

\begin{figure}[t]
    \centering
    \begin{subfigure}{0.48\textwidth}
        \centering
        \includegraphics[width=\linewidth]{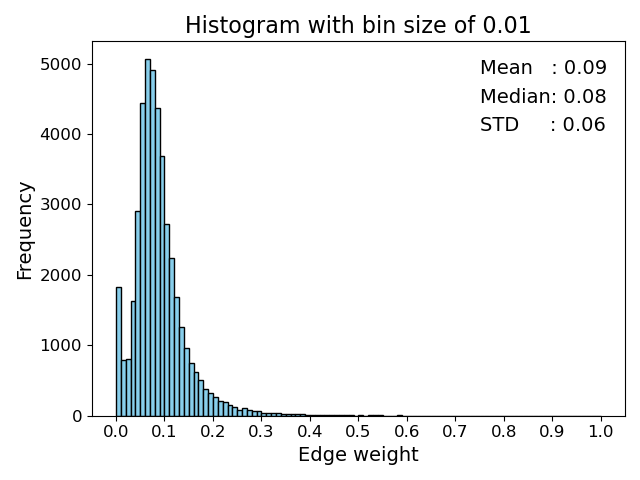}
        \caption{Distribution before cloning}
        \label{fig:woc_dist}
    \end{subfigure} \hfill
    \begin{subfigure}{0.48\textwidth}
        \centering
        \includegraphics[width=\linewidth]{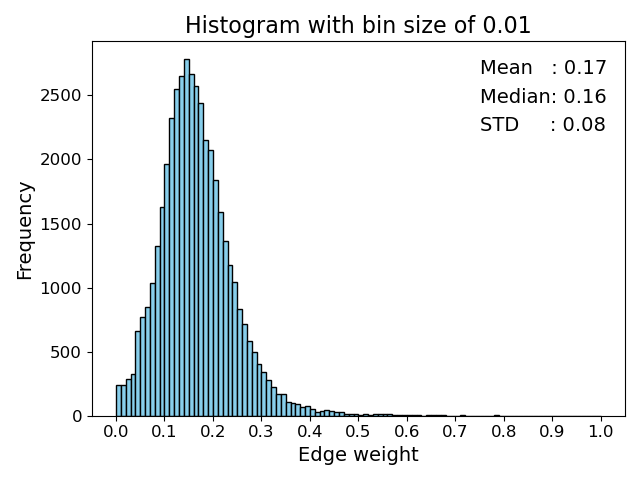}
        \caption{Distribution after cloning}
        \label{fig:wc_dist}
    \end{subfigure}
    \caption{Distribution of network edge weights before and after cloning}
    \label{fig:edge_dist}
\end{figure}

Next, we show the edge weight distribution of the researchers before and after cloning in Figure \ref{fig:edge_dist}. 
Before cloning, our research network contained 296 vertices. After cloning, we added several vertices to accommodate the cloned researchers, resulting in 444 vertices. 
To ensure a fair comparison between the before and after networks, we applied the community refinement step (Section \ref{sec:refine_comm}) to the post-cloning network. 
This step treated the entire network as a single community, merging cloned researchers into a single vertex. When merging vertices, we retained the maximum edge weight (rather than averaging) to preserve the highest topic similarity. 
As shown in Figure \ref{fig:edge_dist}, the network after cloning has an improved similarity edge weight distribution, with reduced skew and increased mean and median values, suggesting that the topical similarity between researchers is more pronounced in the post-cloning network.

\begin{figure}[t]
    \centering
    \includegraphics[width=0.7\linewidth]{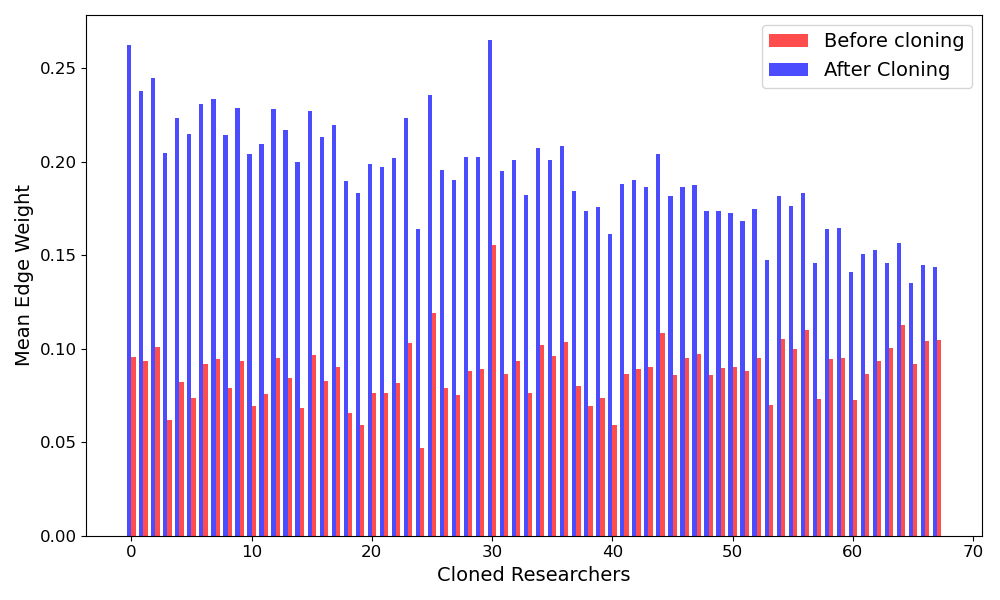}
    \caption{Mean edge weights of the high-output researchers before and after cloning (sorted by difference)}
    \label{fig:norm_edge}
\end{figure}

Figure \ref{fig:norm_edge} shows the mean edge weights of the 68 cloned researchers before and after cloning. 
As with the previous analysis, we ensured comparability between the pre-and post-cloned networks. For all cloned researchers, we observed an increase in mean edge weights, indicating that cloning enhanced the topical distribution of the researchers. 
This suggests that cloning may have more effectively captured diverse topics, leading to higher similarity scores when compared to other researchers, thus showing the effectiveness of the proposed method.

Next we focused on detecting communities in the post-cloned network. 
In the previous analysis, we considered the fully connected network to demonstrate the improved edge weight distribution. 
However, a fully connected network can pose challenges when detecting meaningful communities \cite{noor_nhl}. 
To address this, we fine-tuned the network by applying a threshold of 0.25, which resulted in a network density of 0.1. This threshold allowed us to maintain sufficient edges to preserve relevant connections while preventing the network from being overloaded, thus balancing sparsity and meaningful community structure.

For community detection using the NH-Louvain algorithm, we set a minimum community size threshold of 30. This choice was based on the trade-off between granularity and interpretability. A smaller community size would focus on more specific, high-similarity groups, whereas a larger community size would include a broader set of researchers with less strong topic similarity. 
While the exact threshold is somewhat arbitrary, it depends on the researcher's goal for community granularity: smaller communities for more focused, specialized collaboration or larger communities for broader, more inclusive groupings. 
Table \ref{tab:comm_stat} shows the statistics of the detected communities in the final network after applying the community refinement step.. 
The average community density of 0.64 shows a good level of connectivity between community members, without making the network too dense to affect interpretability.

\begin{table}[t]
\centering
\caption{Summary statistics of the communities in the final research network}
\label{tab:comm_stat}
\begin{tabular}{|l|c|}
\hline
Total Number of Communities     & 30   \\ \hline
Max Community Size          & 28   \\ \hline
Min Community Size          & 2    \\ \hline
Avg. Community Size          & 11.17 \\ \hline
Median Community Size           & 11.5    \\ \hline
Avg. Community Density       & 0.64 \\ \hline
\end{tabular}
\end{table}

With the introduction of cloning, we observed that, after community detection in the post-cloned network, some researchers belonged to more than one community following the community refinement step. 
This outcome aligns with our intuition, as high-output researchers may have diverse topical interests that place them in multiple communities.

%Table \ref{tab:overlap_stat} shows the statistics of overlapping community memberships. 
We found that 29 researchers belong to more than one community. Although clones were created for 68 researchers, after the community refinement step, only 29 of them were placed in multiple communities. 
For the remaining researchers, despite the presence of clones, all of them remained within the same community and were merged into a single vertex during refinement. This suggests that, even though clones were generated, their topical similarity was strong enough to keep them within a single community.

%\begin{table}[t]
%\centering
%\caption{Summary statistics of overlapping researcher memberships}
%\label{tab:overlap_stat}
%\begin{tabular}{|l|c|}
%\hline
%Researchers in multiple communities & 32 \\ \hline
%Max overlapping memberships & 5 \\ \hline
%Min overlapping memberships & 2 \\ \hline
%Median overlapping memberships & 2 \\ \hline
%\end{tabular}
%\end{table}

We also observed that the maximum number of overlapping memberships was 4, indicating that a single researcher was placed in four different communities, reflecting a broad range of topical interests. 
However, we observed the median of 2 overlapping memberships, which suggests that most cloned researchers were placed in just two communities.

\begin{figure}[t]
    \centering
    \includegraphics[width=\linewidth]{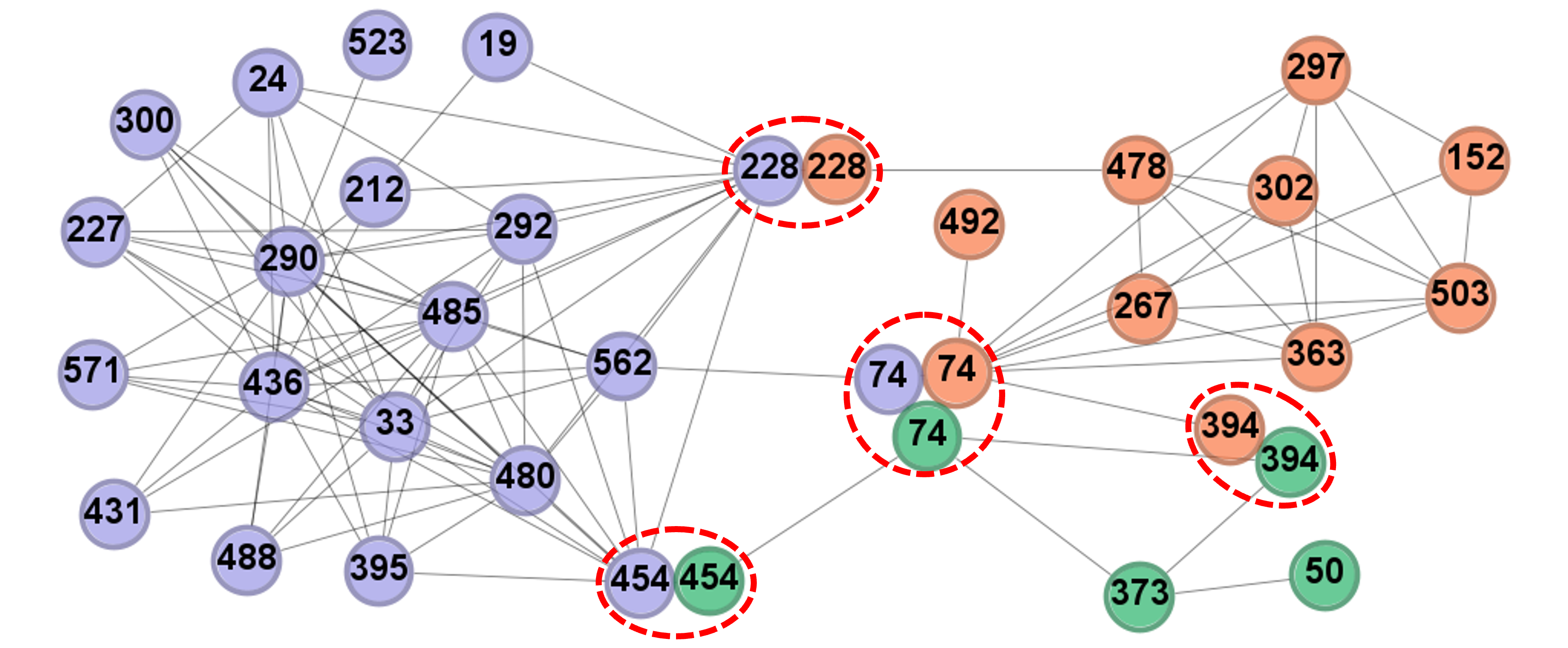}
    \caption{Overlapping communities: dashed circle shows overlapping membership}
    \label{fig:overlapp_ex}
\end{figure}

Figure \ref{fig:overlapp_ex} gives an example subnetwork that shows three communities containing four researchers with overlapping memberships. 
The overlapping nodes are highlighted with dashed circles. Researcher 74 appears in all three communities, while researchers 228, 394, and 454 are part of two communities..
Without cloning, a classic community detection algorithm would have assigned these researchers to a single community based on their most dominant topical similarity. 
However, our cloning approach revealed their diverse topical interests, allowing them to be placed in multiple communities aligned with different research themes.

\begin{figure}[t]
    \centering
    \begin{subfigure}{0.65\textwidth}
        \centering
        \includegraphics[width=\linewidth]{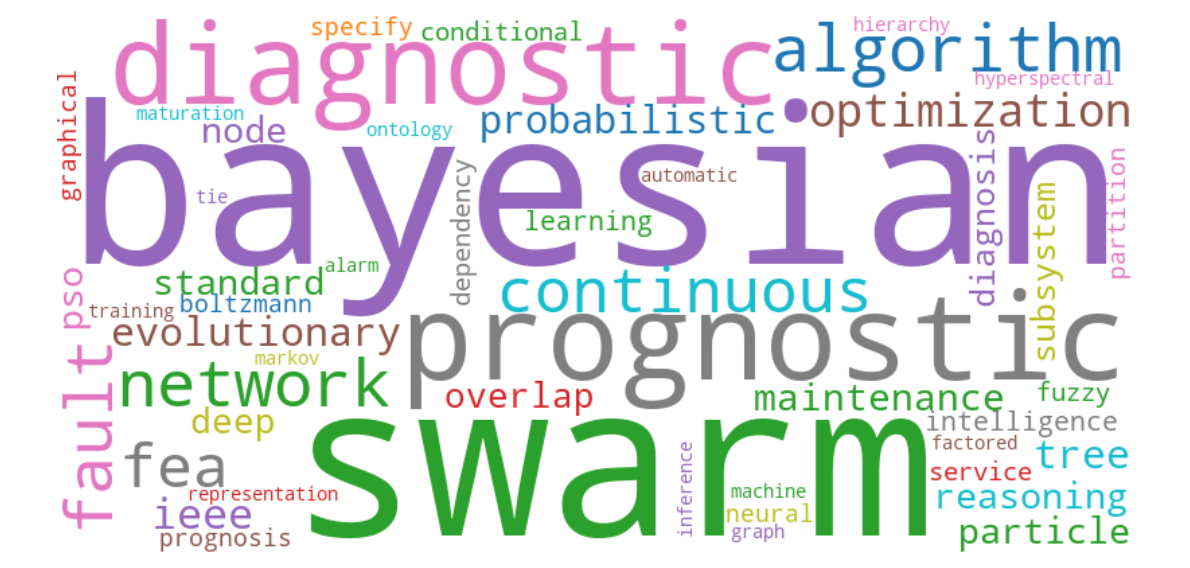}
        \caption{Wordcloud of Researcher 454}
        \label{fig:wc_main}
    \end{subfigure} \hfill
    % \begin{subfigure}{0.48\textwidth}
    %     \centering
    %     \includegraphics[width=\linewidth]{figs/wc_215_0.png}
    %     \caption{Wordcloud of Clone 215-1}
    %     \label{fig:wc_1}
    % \end{subfigure}

    \vspace{0.5em}

    \begin{subfigure}{0.48\textwidth}
        \centering
        \includegraphics[width=\linewidth]{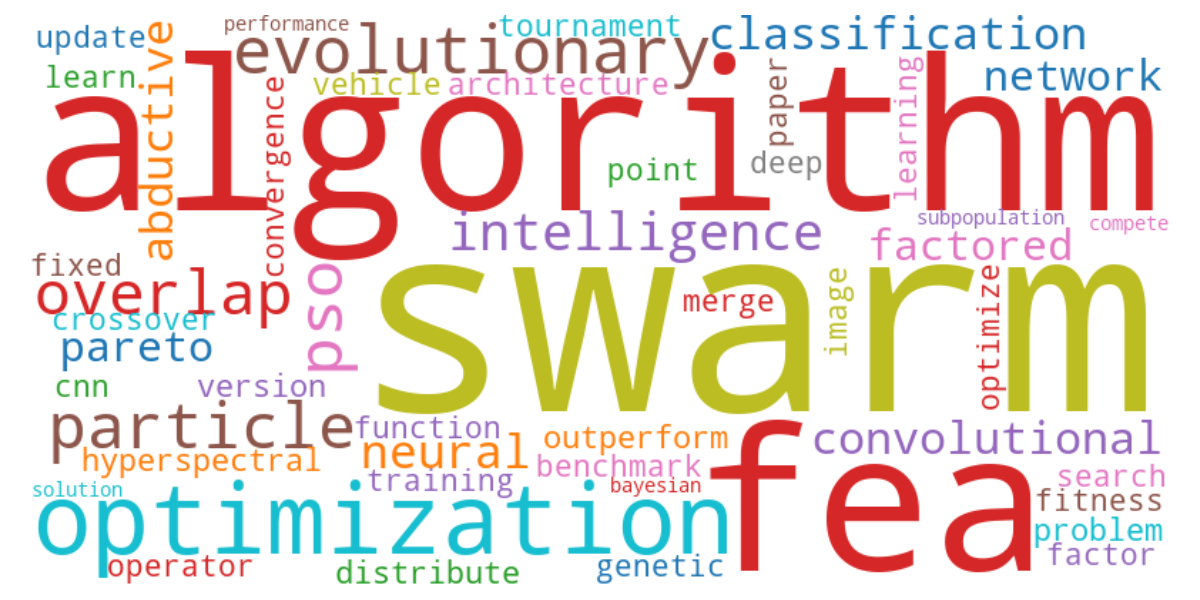}
        \caption{Wordcloud of Clone 454-1}
        \label{fig:wc_2}
    \end{subfigure} \hfill
    \begin{subfigure}{0.48\textwidth}
        \centering
        \includegraphics[width=\linewidth]{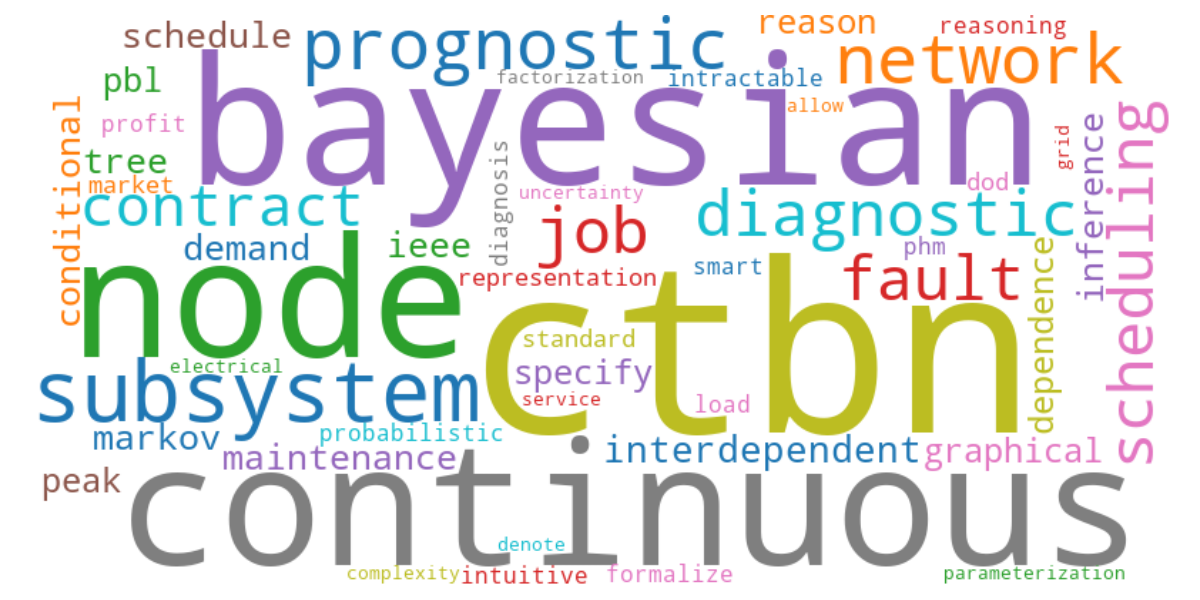}
        \caption{Wordcloud of Clone 454-2}
        \label{fig:wc_3}
    \end{subfigure}

    \caption{Example wordcloud with top 50 words of Researcher 454 with 2 clones}
    \label{fig:wc_215}
\end{figure}

% \begin{figure}[t]
%     \centering

%     \begin{subfigure}{0.5\textwidth}
%         \centering
%         \includegraphics[width=\linewidth]{figs/wc_454_full.png}
%         \caption{Wordcloud of Researcher 454}
%         \label{fig:wc_main}
%     \end{subfigure}

%     \vspace{1em}

%     \begin{subfigure}{0.5\textwidth}
%         \centering
%         \includegraphics[width=\linewidth]{figs/wc_454_comm_13.png}
%         \caption{Wordcloud of Clone 454-1}
%         \label{fig:wc_2}
%     \end{subfigure}

%     \vspace{1em}

%     \begin{subfigure}{0.5\textwidth}
%         \centering
%         \includegraphics[width=\linewidth]{figs/wc_454_comm_9.png}
%         \caption{Wordcloud of Clone 454-2}
%         \label{fig:wc_3}
%     \end{subfigure}

%     \caption{Wordclouds showing the top 50 words for Researcher 454 and its two clones}
%     \label{fig:wc_215}
% \end{figure}

Finally, Figure \ref{fig:wc_215} shows WordCloud examples for the overlapping researcher 454. 
To generate the WordClouds, we first select the researcher’s top 5 topics based on their topic probability distribution. 
Then, for each topic, we multiply the topic probability with the word probability to obtain the top 50 words for visualization. Figure \ref{fig:wc_main} shows the WordCloud for researcher 454 before cloning, where all publications are aggregated. 
The remaining WordClouds correspond to the two clones placed in two different communities. 
As shown, the pre-cloning WordCloud contains terms that combine bio-inspired optimization methods (e.g., factored evolutionary algorithm (FEA) and swarm-based optimization) with probabilistic modeling techniques like continuous-time Bayesian networks (CTBN), reflecting a unified research direction. 
In contrast, the two cloned cases show a separation of these subtopics, with optimization-related terms and Bayesian methods appearing in different clusters. This separation enables the clones to adopt distinct topical focuses and be assigned to different communities.

% As shown, the pre-cloning WordCloud contains a mix of terms from different domains, such as \textit{corrosion, copper, infection, and lung}, making it harder to interpret. 
% After cloning, we see clearer topic separation. For instance, \textit{corrosion} and \textit{copper} appear prominently in Clone 2, while Clone 1 highlights infection-related terms, demonstrating the effectiveness of the cloning approach in capturing distinct topical areas.

Although our method was not designed explicitly as an overlapping community detection approach, the use of cloning naturally resulted in researchers having memberships in multiple communities, creating overlap. 
This raises an interesting question: Could our approach serve as an alternative way to detect overlapping communities in text-based research networks?
Investigating this possibility and comparing it with existing overlapping community detection methods is a promising direction for future work.
In addition, our core idea of handling contribution imbalance through cloning may have broader applicability beyond academic networks. In social networks where a few users dominate content (e.g., online forums, blogs, and product reviews), our method could help identify more balanced community structures that better reflect the contributions of less active users.

\section{Conclusion}
In this paper, we presented an approach and results that explore the potential of topic-based networks for identifying research communities and generating diverse collaboration recommendations. 
Unlike traditional approaches that rely on observable relationships like co-authorship or citations, we focused solely on publication content to build a social network of researchers. 
By leveraging BERTopic and a fine-tuned SciBERT model, we constructed a topic similarity network that identifies connections between researchers across different disciplines.

One of our key contributions was addressing the publication imbalance issue, where researchers with high output tend to be underrepresented in the similarity network. 
To address this, we proposed an effective cloning strategy, clustering their publications to identify less common research areas that would otherwise be overshadowed. 
This allowed researchers to be part of multiple communities, better reflecting their full research landscape and supporting more interdisciplinary recommendations. 
Our evaluation demonstrated that the cloned network structure improves both community coherence and the diversity of potential collaborations.

\subsubsection{Acknowledgments}
This paper is based on work supported, in part, by NSF EPSCoR Cooperative Agreement OIA-2242802.
Any opinions, findings, and conclusions or recommendations expressed in this material are those of the author(s) and do not necessarily reflect the views of the National Science Foundation.

\bibliographystyle{plain}
\bibliography{mybib}

\end{document}